\documentclass[twocolumn]{aastex6}

\usepackage{graphicx}
\usepackage{bm}
\usepackage{psfrag}
\usepackage{mathrsfs}
\usepackage{placeins}
\usepackage{amssymb,amsmath}
\newcommand{\be}{\begin{eqnarray} }
\newcommand{\ee}{ \end{eqnarray} }

\renewcommand{\pasa}{PASA}
\renewcommand{\mnras}{MNRAS}
\renewcommand{\araa}{ARAA}

\renewcommand{\nat}{Nature}
\renewcommand{\be}{\begin{align}}
\renewcommand{\ee}{\end{align}}

\begin{document}
\title{The disturbance of a millisecond pulsar magnetosphere}
\shorttitle{Profile variations in PSR J1643$-$1224}

\shortauthors{R. M. Shannon~et al.}
\author{R. M. Shannon\altaffilmark{1,2}}
\email{ryan.shannon@csiro.au}
\author{L. T. Lentati\altaffilmark{3}}
\author{M.~Kerr\altaffilmark{1}}
\author{M. Bailes\altaffilmark{4}}
\author{N. D. R. Bhat\altaffilmark{2}}
\author{W. A. Coles\altaffilmark{5}}
\author{S. Dai\altaffilmark{1}}
\author{J. Dempsey\altaffilmark{6}}
\author{\\G. Hobbs\altaffilmark{1}}
\author{M. J. Keith\altaffilmark{7}}
\author{P. D. Lasky\altaffilmark{8}}
\author{Y. Levin\altaffilmark{8}}
\author{R. N. Manchester\altaffilmark{1}}
\author{S. Os{\l}owski\altaffilmark{4}}
\author{V. Ravi\altaffilmark{9}}
\author{\\D. J. Reardon\altaffilmark{8,1}}
\author{P. A. Rosado\altaffilmark{4}}
\author{R. Spiewak\altaffilmark{10}}
\author{W. van~Straten\altaffilmark{4}}
\author{L. Toomey\altaffilmark{1}}
\author{J.-B. Wang\altaffilmark{11}}
\author{L. Wen\altaffilmark{12}}
\author{\\X.-P. You\altaffilmark{13}}
\author{X.-J. Zhu\altaffilmark{12}}
\altaffiltext{1}{CSIRO Astronomy and Space Science, Australia Telescope National Facility, Box 76, Epping NSW 1710, Australia}
\altaffiltext{2}{International Centre for Radio Astronomy Research, Curtin University, Bentley WA 6102,  Australia}
\altaffiltext{3}{Astrophysics Group, Cavendish Laboratory, JJ Thomson Avenue, Cambridge CB3 0HE, UK} 
\altaffiltext{4}{Centre for Astrophysics and Supercomputing, Swinburne University of Technology, P.O. Box 218, Hawthorn, Victoria 3122, Australia}
\altaffiltext{5}{Department of Electrical and Computer Engineering, University of California at San Diego, La Jolla, CA 92093, USA}
\altaffiltext{6}{CSIRO Information Management \& Technology, Box 225, Dickson ACT 2602 }
\altaffiltext{7}{Jodrell Bank Centre for Astrophysics, University of Manchester, M13 9PL, UK.}
\altaffiltext{8}{Monash Centre for Astrophysics, School of Physics and Astronomy, Monash University, VIC 3800, Australia}
\altaffiltext{9}{Cahill Center for Astronomy and Astrophysics, MC 249-17, California Institute of Technology, Pasadena, CA 91125, USA}
\altaffiltext{10}{Department of Physics, University of Wisconsin-Milwaukee, Box 413, Milwaukee, WI 53201, USA}
\altaffiltext{11}{Xinjiang Astronomical Observatory, Chinese Academy of Sciences, 150 Science 1-Street, Urumqi, Xinjiang 830011, China}
\altaffiltext{12}{School of Physics, University of Western Australia, Crawley, WA 6009, Australia}
\altaffiltext{13}{School of Physical Science and Technology, Southwest University, Chongqing, 400715, China}




\begin{abstract}
Pulsar timing has enabled some of the strongest tests of fundamental physics. 
Central to the technique is the assumption that the detected radio pulses can be used  to accurately measure the rotation of the pulsar. 
Here we report on a broad-band variation in the pulse profile of the millisecond pulsar J1643$-$1224.  
A new component of emission  suddenly appears in the pulse profile, decays over $4$~months, and results in a permanently modified pulse shape. 
Profile variations such as these may be the origin of  timing noise observed in other millisecond pulsars. 
The sensitivity of pulsar-timing observations to gravitational radiation can be increased by accounting for this variability. 

\end{abstract}


\section{Introduction}

\begin{figure*}[!ht]
\begin{center}
\begin{tabular}{cc}
\includegraphics[scale=0.8]{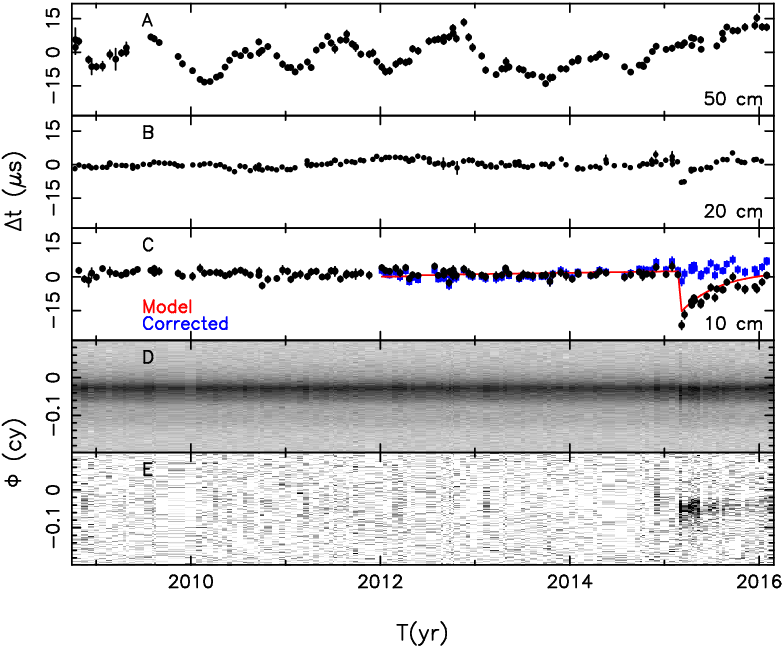}  & \hspace{0.1in}  \includegraphics[scale=0.8]{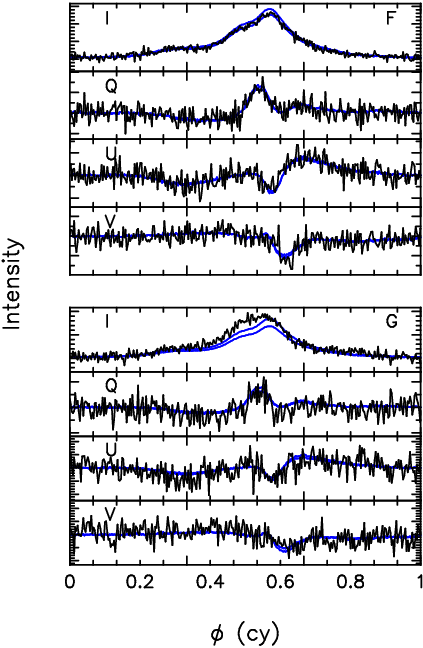} \\
\end{tabular}
\end{center}
\caption{ 
 \label{fig:timing_profile}  
Timing and profile variations for PSR~J1643$-$1224 in our most recent observations.    In the residual TOAs (panels A-C), we have removed a quadratic DM variation assuming cold-plasma dispersion frequency dependence.     A:  Residual TOAs in the 50-cm band.  B:  Residual TOAs in the 20-cm  band. C: Residual TOAs (black points) in the $10$-cm  band.  The red line shows the TOA perturbation introduced when using a standard template to form TOAs from the maximum {\em a-posteriori} model pulse profiles including the shape variations. This curve has been derived from the preferred model for the profile variations (Model 7 in Table \ref{tab:evidence}) derived from the profile domain analysis (see Section \ref{sec:profiledomain}).   This model was chosen because it had an evidence comparable to the highest-evidence model (Model 9), but fewer parameters. 
   The blue points then show the residual TOAs after subtracting the red line.     D:  Integrated pulse profiles for the 10-cm observations. E:  Residual profiles in the $10$-cm band, assuming a non-varying integrated profile and correcting for pulse amplitude using polarized flux.  F:  10-cm pulse profile (black lines)  at epoch prior to change (21~Feb~2015) .  The blue lines show the ($1\sigma$) maximum-likelihood range for the predicted amplitude of the pulse profile as measured from the polarized flux.   G.  10-cm pulse profile at epoch after  change (7~March~2015).       }
\end{figure*}

The measurement of arrival times (TOAs) of pulses from  radio pulsars has enabled physical tests impossible in terrestrial laboratories or elsewhere in the Universe,  including studies of nuclear equations of state \cite[][]{2010Natur.467.1081D} and strong-field tests of the general theory of relativity \cite[][]{2006Sci...314...97K}.
Another goal of precision timing is the direct detection of gravitational radiation.  
By monitoring TOAs from an array of spin-stable millisecond pulsars  (MSPs) for months to decades, it is likely possible to detect the distortions in TOAs induced by gravitational waves travelling through the solar neighbourhood, and to distinguish their quadropolar signature from contaminating noise sources \cite[][]{1990ApJ...361..300F,2010CQGra..27h4013H,2016MNRAS.455.4339T}.  
Two properties of pulsars are central to the power of the  pulsar-timing technique. 
Firstly, pulsars are rotationally stable, with the spin of the pulsar well modelled deterministically  and every rotation of the pulsar accounted for over year to decades-length observing spans.
Second is the assumption is that the pulsar radio emission -- located in the diffuse pulsar magnetosphere tens to thousands of kilometres above the neutron-star surface -- is anchored to the star.

Although individual pulses show markedly different morphology  \cite[e.g.,][]{cs2010}, a third assumption is that the longitude-resolved average emission profile converges towards a stable shape after many rotations of the pulsar.
It has been known for decades that many slowly spinning pulsars have profiles that switch between distinct emission modes on time scales of minutes to hours (referred to as mode-switching).   
More recently, \cite{2010Sci...329..408L} identified subtle pulse profile variations in long-term observations of young pulsars, with the pulsars switching modes on month to decade time scales, and the state switches correlated with discrete changes in spin down. 

Pulse-shape variations can also be caused by external factors. 
For example,  precession of the pulsar spin axis has been observed for pulsars in the most compact, relativistic orbits \cite[][]{1989ApJ...347.1030W}.
  As the orientation of the axis changes with respect to our line of sight,   regions of different radio luminosity  are beamed toward the earth.  
Interstellar propagation effects are also predicted to cause pulse-profile variations \cite[][]{cs2010}. 
  As the pulsar and Earth move, the line of sight samples a different column of interstellar material, which can potentially cause variable pulse broadening. 

There is only modest evidence for pulse-shape variations in millisecond pulsars central to pulsar timing array observations.  
Claims of epoch-to-epoch mode-switching in the millisecond pulsar J1022$+$1001 \cite[][]{1999ApJ...520..324K} have been disputed.   
These profile variations  can instead be attributed to  instrumental-calibration errors  \cite[][]{2004MNRAS.355..941H,2013ApJS..204...13V} and  spectral evolution of the profile \cite[][]{2015MNRAS.449.3223D,2003AA...407.1085R}. 
Other well-monitored MSPs have shown no evidence for pulse-profile evolution over decades-long time scales \cite[][]{2013CQGra..30p5019S}. 




\section{Observations}
\label{sec:obs}

As part of the Parkes Pulsar Timing Array project \cite[PPTA,][]{2013PASA...30...17M}, we have been observing PSR J1643$-$1224 with the $64$-m Parkes telescope at a $\approx 3$-week cadence since $2003$.
 At each epoch,  the pulsar is normally  observed in three frequency bands:  at a frequency close to $1400$~MHz using either the central feed of the 20-cm multibeam  system or  the H$-$OH receiver, and simultaneously at frequencies close to $3100$~MHz and $700$ MHz using the dual-band 10cm/50cm system.  
 Observations have been made with a range of backends, but the observations central to the results  here were made with the mark-3 and 4 Parkes digital filterbank systems (PDFB3/4) and the CASPER-Parkes-Swinburne recorder (CASPSR). 
 Since 2014 May, the PDFB3 backend has been unavailable because of hardware failure. 
 Since 2015 June, the $50$-cm band has been  affected by radio-frequency interference from  a 4G base station $10$-km north of the telescope  in nearby Alectown 

The backend systems use fundamentally different processing techniques to process the voltage time series data: the PDFB systems employ digital polyphase filterbanks produced on field-programmable gate arrays and CASPSR employ coherent dedispersion on  graphics-processing units.  
Both systems use online folding to produce average spectra (in four pseudo-Stokes parameters) that sample pulse phase and average over time.     

The observations were calibrated and analyzed using standard pulsar processing techniques as implemented in the code {\sc  psrchive} \cite[][]{2004PASA...21..302H} within the PPTA data-reduction pipeline  \cite[][]{2013PASA...30...17M}.   
In brief, median filters were applied to the spectra to mitigate the effects of radio-frequency interference and 
variable differential gain and phase of the system were measured using regular observations of a noise-diode signal, injected 45$^\circ$ from the linear feeds.
The flux-density scale of the observations was determined by relating the strength of the noise-diode signal  to regular observations of the radio galaxy Hydra~A.  
For observations made with the 20-cm multibeam system, we also corrected for  cross-coupling in the feed.

 \label{sec:timing}

 \section{Timing Analysis}

Evidence for changes in pulse shape were first identified in timing analyses.   The TOAs were produced by cross-correlating the calibrated Stokes-$I$ profiles with an analytic template, 
  constructed from the average of observations  from  2008-2014 \cite[][]{2015MNRAS.449.3223D}. 
   Analysis of the TOAs  was conducted using the {\sc tempo2}  \cite[][]{2006MNRAS.372.1549E} and {\sc temponest} \cite[][]{2014MNRAS.437.3004L} codes.
TOAs were referred to the solar system barycentre using the DE421 ephemeris published by the Jet Propulsion Laboratory and the 2015 realization of terrestrial time published by the International Bureau of Weights and Measures.  
 
 In Figure \ref{fig:timing_profile}, we show the maximum-likelihood residual TOAs obtained for PSR~J1643$-$1224, individually in the $50$~cm (panel A), $20$-cm (panel B), and $10$-cm bands (panel C). 
The variations in interstellar dispersion were modelled  with a quadratic (in time) polynomial, using only  10-cm and 20-cm observations. 
 This only removes the lowest fluctuation frequency dispersion-measure (DM) variations, but is sufficient for our analysis. 
 
Relative to  this model, we find that the $50$-cm observations show annual TOA variations inconsistent with the $10$-cm and $20$-cm observations.  
The level of  signal in these variations scale $\propto \lambda^{4}$, suggesting that they are associated with scattering in the interstellar medium \cite[][]{iptanoise}.
They are likely the result of the Earth-pulsar line of sight resampling the same region of the ISM for this low-velocity  pulsar \cite[$v_\perp \approx$ 25 km~s$^{-1}$,][]{2016MNRAS.455.1751R}.
  
  The largest departure of the residuals occurred since $2015$~March, predominantly in the shortest-wavelength ($10$-cm) band.
 Between observations of $2015$~February~21 and $2015$~March~7   the  apparent TOAs in this band arrived  $\approx 25~\mu$s earlier. 
  In the 20-cm band, the TOAs shifted by $\approx 10~\mu$s between 2015~19~February and 6~March. 
    In contrast,  no significant arrival-time variations were observed at $50$~cm, meaning that these variations are unlikely to be caused by changes in the ISM.
       Since then, the  offset has decreased in amplitude, with the pulses now arriving $\approx~5~\mu$s earlier than the long term pre-2015 level.  
    
These variations are unlikely to be instrumental in origin. 
In the 20-cm band, the TOA variations are common to two backends (PDFB4 and CASPSR). 
In the 10-cm band, other pulsars observed with the same backend (PDFB4) in the same firmware configuration do not show this offset, with a limit on the change of arrival time of $< 100$~ns for the most stable pulsar in the sample, PSR~J1909$-$3744 \cite[][]{2015Sci...349..1522S}. 
The lack of similar variations in other pulsars excludes other telescope-dependent (but backend-independent) origins, including an incorrect time or frequency reference at the observatory. 
In particular, PSR~J1643$-$1224 is less susceptable to polarization calibration errors than many other pulsars in the PPTA sample \cite[][]{2013ApJS..204...13V}.

\begin{figure}
\begin{center}
 \includegraphics[scale=0.6]{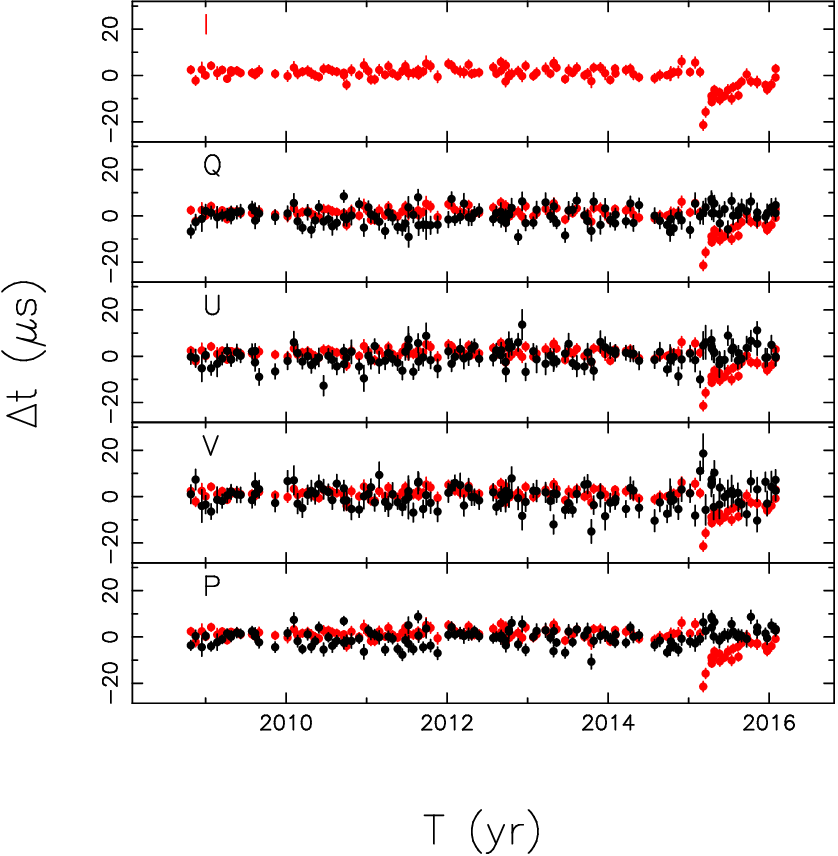}  \end{center}
\caption{Maximum-likelihood residual TOAs for PSR~J1643$-$1224, timing in Stokes parameters $I$, $Q$, $U$, and $V$ as listed on the plot. 
  The bottom panel (labelled $P$) shows the average residuals of the polarized components, weighted by their uncertainties.  The Stokes-$I$ TOAs are plotted in red in all panels.      }
\label{fig:toas_pol}
\end{figure}

The TOAs in other Stokes components do not show the same sudden variation or recovery.
In Figure \ref{fig:toas_pol}, we show the TOAs determined from timing individual Stokes parameters.  
None showed the magnitude of variation observed in Stokes $I$.
In contrast, {\em bona-fide} arrival-time variations would affect all Stokes components equally.  
A pulse-shape change could manifest  differently in  the four Stokes components, and would also introduce observed apparent arrival-time variation.

\section{Pulse-shape variations}
\label{sec:shape}

Identifying and characterizing  changes in pulse shape are complicated by the unknown {\em a-priori} pulse arrival times and flux densities. 
Pulse flux densities show marked variations from epoch to epoch as the received strength is modulated by diffractive and refractive scintillation of the pulsar radiation in the interstellar medium \cite[][]{1990ARAA..28..561R}.
The pulse arrival time can be predicted from a model, but will be incorrect if an unmodeled or poorly modeled process is present in the observations. 
When pulse profiles are shifted and scaled for these unknown parameters, much of the shape variations can be absorbed into these two parameters.

Because the polarized emission of PSR~J1643$-$1224 is nearly unaffected by the shape change, it can be used to measure the expected amplitude and phase of the total intensity pulse profile. 
Prior to the March~2015 event, we assumed that the pulse shape was constant and then extrapolate the long term model after the event. 

An amplitude estimator for Stokes $I$ was calculated by averaging the individual amplitude estimates from Stokes $Q$, $U$, and $V$.   
This assumes that all Stokes parameters experience the same scintillation amplitude variations, justified as there is no evidence for birefringence of scintillations \cite[][]{1984ApJ...284..126S}.
The average pulse profiles for the recent 10-cm observations of J1643$-$1224 are displayed in panel D of Figure \ref{fig:timing_profile}.  
The residual of these profiles relative to a model formed from all the pre-March 2015 observations is displayed in panel E of Figure \ref{fig:timing_profile}.  
Prior to March 2015, the on-pulse residuals  (phase  $\phi =-0.2$ to $\phi\approx 0.$)  show nearly noise-like  variations.
Significant excess signal is observed on the leading edge of the pulse profile after March 2015, commensal with the sudden change in residuals plotted in the bottom panel of the Figure. 
The pulse profiles from the epochs immediately before (panel F) and after (panel G) highlight the emergence of the new component in the pulse profile. 
Neither the residual TOAs nor the residual profiles return to the pre-event levels, with 10-cm and 20-cm profiles showing permanent excess power on the leading edge of the profile, resulting in the post-event TOAs showing an apparent permanent offset of  $-5~\mu$s.

\vspace{0.6in}
\section{Profile-domain timing}

\label{sec:profiledomain}


We  analysed our observations using a recently developed profile-domain timing methodology that  implements pulsar-timing analyses directly on  pulse profiles \cite[][]{2015MNRAS.447.2159L,2015MNRAS.454.1058L}.
The algorithm incorporates a  likelihood function that simultaneously describes deterministic and stochastic contributions to both the pulse profile and the TOAs.  
It uses Bayesian methodology to sample the posterior distribution,  enabling  robust parameter estimation and  marginalization over  other nuisance parameters.
   It also enables the quantitative comparison of families of models through the use of evidence (the  integral of the likelihood weighted by the prior).  
When evidences are compared,  differences in the $\log$ evidence of $\ge 3$ delineate a clear model preference with a probability of $\ga 95\%$.  For differences in the $\log$ evidence  of $ \Delta \log E < 3$, the simpler model is preferred.
The method naturally accounts for the covariance between shape and timing variations and can therefore better break the degeneracy between the two. 
The models considered are listed in Table \ref{tab:evidence}.

Non-stationary profile variations were parameterized using a shapelet basis \cite[][]{2003MNRAS.338...35R},  comprising a series of  Hermite polynomials multiplied by a Gaussian.
This basis provides a series of functions that compactly model pulse profiles and profile variations,  using only two  non-linear parameters (the centre of the profile variation and its  width) for the entire basis.

We considered both permanent and transient changes to the pulse profile. 
For the transient changes, we compared models where the amplitude of the additional components were assumed to exponentially decay after an initial epoch, or to rise and fall like a Gaussian. 
These models are listed in the second (transient) and third (permanent) columns of Table \ref{tab:evidence}.   

In  conjunction with the profile modelling, we also searched for timing noise, considering both stationary timing noise modelled using a power-law process and predicted to be present in many MSPs \cite[][]{sc2010} (fourth column of Table \ref{tab:evidence}), and nonstationary timing noise which could be described using the shapelet basis \cite[fifth column, see also ][]{iptanoise}.

Motivated by the shape of the perturbation as seen in the TOAs, we also considered TOA perturbations that, after an initial epoch, either exponentially decayed (sixth column of Table \ref{tab:evidence}), similar to the glitch decays observed in many pulsars \cite[][]{2013MNRAS.429..688Y}, or resulted in a permanent offset (seventh column). 
A permanent offset would be indicative of an unmodelled transient change in the spin state. 
When the same type of TOA perturbations and shape variations (i.e., permanent or exponential) were modelled simultaneously, the onset epoch and decay timescales were modelled as being common to both.  
While   changes in shape and TOA may not necessarily share the same physical timescale, we used the common model parameters to ensure preference for either shifts or shape changes could be determined from the posterior parameter estimates.


In order to reduce computation requirements, only  the most recent 4~yr of observation (see Figure \ref{fig:timing_profile}) were considered, we analyzed the $10$-cm and $20$-cm data sets independently, and we did not consider the $50$~cm observations. 
We found that the data strongly support a transient change in the pulse profile, with exponential models (Models 5-12 in Table \ref{tab:evidence}) favored over those with Gaussian variation (Model 4).  The shape variations can be adequately modelled using only two shapelet components, with the evidence decreasing as additional components are added to the model.  

A non-stationary timing noise origin for the TOA variations (Model 2)  is strongly disfavored.  In addition,  we find that the data support a model where no TOA shifts are required in addition to the shape variation (Model 7).  
 
Our Stokes-$I$ observations cannot distinguish between a permanent change in the pulse profile or a permanent offset in the TOAs (Models 6 and 7); however the lack of an offset in the arrival times of the other Stokes components suggest that the profile has changed shape. 
In this case, we limit the shift in arrival times to be $< 1 \mu$s.
After accounting for the pulse profile variations (and possible permanent offset) we find no evidence for excess noise in the TOAs.
Evidence comparisons of models in the 20 cm band result in the same conclusions.

\begin{figure}
\begin{center}
 \includegraphics[scale=0.4]{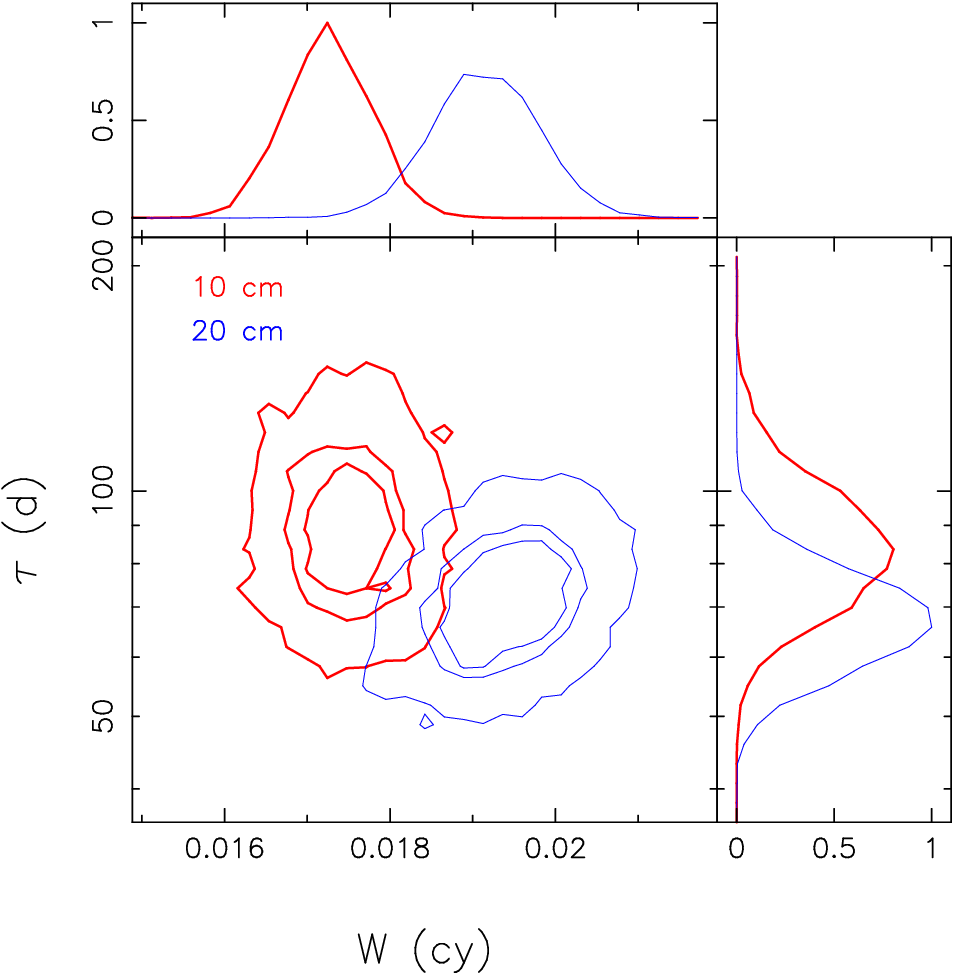}   
 \end{center}
\caption{
Marginalized posterior distributions for decay time ($\tau$) and width ($W$), as measured via profile-domain timing of our 10-cm (red) and 20-cm (blue) data.}
\label{fig:post_distro}
\end{figure}

In Figure \ref{fig:post_distro}, we show the posterior distribution for the shape variations in both the $10$-cm and $20$-cm bands.
We find that the components decay with a common time scale $\tau$, but the 20-cm component is broader than that observed at $10$~cm. 

After accounting for the shape variations results, the TOAs are less biased, extrapolating from the long-term model.
In panel C of Figure \ref{fig:timing_profile}, we show the perturbations associated with one of the favored pulse-shape variation models  (Model 7)  as a red line and a maximum likelihood representation of the corrected TOAs.
The blue points show the residuals corrected by the model and do not show any variation or permanent offset. 

\begin{deluxetable}{crcccccr}
\tabletypesize{\scriptsize}
\tablecaption{Models and evidences \label{tab:evidence}}
\tablehead{  & \multicolumn{6}{c}{Model}   \\
\cline{2-7} 
 & \multicolumn{2}{c}{Shape variations} & \multicolumn{4}{c}{Timing noise} & \\ 
 \cline{2-3} \cline{4-7}
 Model &  Trans. & Perm. & Stat. & NS & Trans. & Perm. &  \colhead{$\Delta \log E$} \\
  & (\#) &  (\#) & & & }
 \startdata
1   &  0     & 0 	 			& $\checkmark$		& $\times$ & $\checkmark$  & $\checkmark$          &  -84.2 \\ 
 2   &  0  & 0 		&  $\checkmark$				& $\checkmark$ & $\checkmark$  & $\checkmark$  &  -88.0 \\ 
  3      &  0     & 2 			&  $\checkmark$		& $\times$ & $\checkmark$  & $\checkmark$          &  -33.4 \\ 
 4      &  2g & 0 				&  $\checkmark$  		& $\times$ &  $\checkmark$  	& $\checkmark$          &  -25.4 \\
5      &  2e    & 0 			&  $\checkmark$		& $\times$ & $\checkmark$  & $\times$              &  -9.7  \\ 
 {\bf 6} & {\bf 2e}     	& {\bf 0}    &  $\bm{\checkmark}$  	& ${\bm \times}$ &$\bm{\checkmark}$  & $\bm{\checkmark}$          &  {\bf -2.0}  \\ 
{\bf 7} & {\bf  2e}     & {\bf 2}	& $\bm{\checkmark}$ 	& ${\bm \times}$& $\bm{\times}$      & $\bm{\times}$              &  {\bf -2.7}  \\ 
{\bf 8} & {\bf  2e}     & {\bf 2} 	&  $\bm{\checkmark}$ 	& ${\bm \times}$& $\bm{\checkmark}$  & $\bm{\checkmark}$          &  {\bf -1.3}  \\ 
{\bf 9} &  {\bf 2e}     & {\bf 2} 	&  $\bm{\checkmark}$ 	& ${\bm \times}$& $\bm{\times}$      & $\bm{\checkmark}$          &   {\bf 0}   \\ 
{\bf 10} & {\bf 2e} & {\bf 2} 		& $\bm{\checkmark}$ 	& $\times$& $\checkmark$ & $\times$ & {\bf -2.5} \\
11      &  3e     & 2 			&  $\checkmark$		& $\times$& $\checkmark$  & $\checkmark$          &  -4.3  \\  
12      &  4e    & 2 			&  $\checkmark$		& $\times$&  $\checkmark $ & $\checkmark$          &  -1.8  \\ 
\enddata

\tablecomments{Models in boldface are considered plausible.    For the transient shape variations (column 2), the numerical value is the number of components used and the letter represents the envelope for the variations, with  $g$ indicating a Gaussian envelope and $e$ an exponential envelope.      Relative evidences are measured relative to the Model 9, which has the highest evidence. }
\end{deluxetable}

\section{Discussion and conclusions}
\label{sec:discuss}

Secular shape variations  have not hitherto been observed in a millisecond pulsar.
In state-changing young pulsars, pulse shapes change between discrete modes \cite[][]{2010Sci...329..408L},  with strong correlations identified between change in pulse shape and rotation state.
The  shape variations observed in PSR~J1643$-$1224 are most reminiscent of those observed in the young pulsar B0736$-$40 \cite[][]{2011MNRAS.415..251K},  in which a new component suddenly appeared and gradually drifted towards the central emission region over a few years of observation.  
In both PSRs~J1643$-$1224 and B0736$-$40, the new components are unpolarized. 
The lack of a change in spin down (manifested as an offset in the TOAs) is in contrast to what is observed in other pulsars like PSR B0736$-$40.
A mean change in the spin down rate of $\langle\Delta \dot{\nu}\rangle$ over a time scale $T$ would result in a permanent offset in pulse phase  $\Delta \Phi \approx \langle\Delta \dot{\nu}\rangle  T^2/2$. 
Because the shift in the arrival times is limited to  $ < 1~\mu$s  after $100$~d, the change in spin down is $|\Delta \dot{\nu}| \lesssim 2 \times 10^{-20}$~s$^{-2}$, which is a factor $10^{4}$ smaller than the long-term spin down of the pulsar \cite[][]{2016MNRAS.455.1751R} and two orders of magnitude  fractionally smaller than the correlated spin down variations observed in other pulsars.

In PSR~B0736$-$40, the profile variation was attributed to the impact of an asteroid on the neutron-star magnteosphere \cite[][]{2014ApJ...780L..31B}.  
It is unlikely that an extrinsic interaction is the cause of the profile variation observed here.  
The magnetospheres of MPSs are factors of $10$-$100$ smaller than that of  slower-rotating pulsars like B0736$-$40, making asteroidal intrusion a less likely explanation for the profile variation \cite[][]{2008ApJ...682.1152C}. 
Furthermore, compared to solitary pulsars, binary systems such as PSR~J1643$-$1224 potentially harbour less massive and wider disks  \cite[][]{2013ApJ...766....5S}  with fewer mechanisms to dynamically inject material into the pulsar magnetosphere. 


For gravitational wave detection experiments,  it is essential to search for profile variability, and  correct for it, if it exists. 
As an example, we placed a limit on the gravitational-wave background  in the  four most recent years of 10cm PDFB4 observations of PSR~J1643$-$1224   using both standard methods and profile-domain timing.
Limits placed in the profile domain follow methods presented in  \cite[][]{2015MNRAS.454.1058L} and limits in the TOA domain follow methods presented in \cite[][]{2015Sci...349..1522S}. 
We assumed a background with a strain spectrum of the form $h_c(f)=A_{c, {\rm yr}} f^{-2/3}$, where $f$ is the frequency measured in cycles per year and
 $A_{c, {\rm yr}}$ is the amplitude of the background. This is the form predicted for a background produced by binary supermassive black holes that are radiating all of their energy in gravitational radiation. 
When neglecting to account for the profile variations, we find  $A_{c, {\rm yr}} < 2\times 10^{-12}$. 
When accounting for the variability, the limit improved to  $A_{c, {\rm yr} }< 9 \times 10^{-14}$. 
Improvements to other pulsars will depend on the properties and signal-to-noise ratio of the profile variations and the spectral color that the variations induce in the TOAs.

It is possible that shape variations induce the timing noise observed in many MSPs. 
Red timing noise has been identified as being common in MSPs \cite[][]{2013PASA...30...17M,2015ApJ...813...65T,2016MNRAS.457.4421C,2016MNRAS.455.1751R}.  
In contrast to observations of timing noise in young pulsars and  prediction for MSPs (and in observed in a minority)   \cite[][]{sc2010}, much of the  observed noise in MSPs has a relatively shallow or non-power-law  spectral shape  \cite[][]{2015ApJ...813...65T,2016MNRAS.455.1751R}.  
If this noise is associated with shape variations, it may be possible to mitigate it.  
The ability to correct for shape variations will depend on the signal-to-noise ratio of the variations.
High-fidelity observations from large-aperture telescopes like the Five-hundred-metre  aperture spherical telescope  \cite[][]{2011IJMPD..20..989N,2014arXiv1407.0435H}  and the Square Kilometre Array \cite[][]{2015aska.confE..37J} will be particularly amenable to the identification and correction of these magnetospheric disturbances. 

\acknowledgements
The Parkes radio telescope is part of the Australia Telescope which is funded by the Commonwealth of Australia for operation as a National Facility managed by the Commonwealth Science and Industrial Research Organization (CSIRO). The Parkes Pulsar Timing Array Project project was initiated with support  from an Australian Research Council (ARC) Federation Fellowship (FF0348478) to RNM and from the CSIRO under that fellowship program. The PPTA project has also received support from the ARC through Discovery Project grants DP0985272 and DP140102578.  NDRB acknowledge support form a Curtin University research fellowship.   GH and YL are recipients of ARC Future Fellowships (respectively, FT120100595 and FT110100384).  SO is supported by the Alexander von Humboldt Foundation. RMS acknowledges travel support from the CSIRO through a John Philip award for excellence in early career research. 


\begin{thebibliography}{}
\expandafter\ifx\csname natexlab\endcsname\relax\def\natexlab#1{#1}\fi

\bibitem[{{Arzoumanian} {et~al.}(2015){Arzoumanian}, {Brazier},
  {Burke-Spolaor}, {Chamberlin}, {Chatterjee}, {Christy}, {Cordes}, {Cornish},
  {Crowter}, {Demorest}, {Dolch}, {Ellis}, {Ferdman}, {Fonseca},
  {Garver-Daniels}, {Gonzalez}, {Jenet}, {Jones}, {Jones}, {Kaspi}, {Koop},
  {Lam}, {Lazio}, {Levin}, {Lommen}, {Lorimer}, {Luo}, {Lynch}, {Madison},
  {McLaughlin}, {McWilliams}, {Nice}, {Palliyaguru}, {Pennucci}, {Ransom},
  {Siemens}, {Stairs}, {Stinebring}, {Stovall}, {Swiggum}, {Vallisneri}, {van
  Haasteren}, {Wang}, \& {Zhu}}]{2015ApJ...813...65T}
{Arzoumanian}, Z., {Brazier}, A., {Burke-Spolaor}, S., {et~al.} 2015, \apj,
  813, 65

\bibitem[{{Brook} {et~al.}(2014){Brook}, {Karastergiou}, {Buchner}, {Roberts},
  {Keith}, {Johnston}, \& {Shannon}}]{2014ApJ...780L..31B}
{Brook}, P.~R., {Karastergiou}, A., {Buchner}, S., {et~al.} 2014, \apjl, 780,
  L31

\bibitem[{{Caballero} {et~al.}(2016){Caballero}, {Lee}, {Lentati}, {Desvignes},
  {Champion}, {Verbiest}, {Janssen}, {Stappers}, {Kramer}, {Lazarus},
  {Possenti}, {Tiburzi}, {Perrodin}, {Os{\l}owski}, {Babak}, {Bassa}, {Brem},
  {Burgay}, {Cognard}, {Gair}, {Graikou}, {Guillemot}, {Hessels},
  {Karuppusamy}, {Lassus}, {Liu}, {McKee}, {Mingarelli}, {Petiteau}, {Purver},
  {Rosado}, {Sanidas}, {Sesana}, {Shaifullah}, {Smits}, {Taylor}, {Theureau},
  {van Haasteren}, \& {Vecchio}}]{2016MNRAS.457.4421C}
{Caballero}, R.~N., {Lee}, K.~J., {Lentati}, L., {et~al.} 2016, \mnras, 457,
  4421

\bibitem[{{Cordes} \& {Shannon}(2008)}]{2008ApJ...682.1152C}
{Cordes}, J.~M., \& {Shannon}, R.~M. 2008, \apj, 682, 1152

\bibitem[{{Cordes} \& {Shannon}(2010)}]{cs2010}
---. 2010, arXiv:1010.3785

\bibitem[{{Dai} {et~al.}(2015){Dai}, {Hobbs}, {Manchester}, {Kerr}, {Shannon},
  {van Straten}, {Mata}, {Bailes}, {Bhat}, {Burke-Spolaor}, {Coles},
  {Johnston}, {Keith}, {Levin}, {Os{\l}owski}, {Reardon}, {Ravi}, {Sarkissian},
  {Tiburzi}, {Toomey}, {Wang}, {Wang}, {Wen}, {Xu}, {Yan}, \&
  {Zhu}}]{2015MNRAS.449.3223D}
{Dai}, S., {Hobbs}, G., {Manchester}, R.~N., {et~al.} 2015, \mnras, 449, 3223

\bibitem[{{Demorest} {et~al.}(2010){Demorest}, {Pennucci}, {Ransom}, {Roberts},
  \& {Hessels}}]{2010Natur.467.1081D}
{Demorest}, P.~B., {Pennucci}, T., {Ransom}, S.~M., {Roberts}, M.~S.~E., \&
  {Hessels}, J.~W.~T. 2010, \nat, 467, 1081

\bibitem[{{Edwards} {et~al.}(2006){Edwards}, {Hobbs}, \&
  {Manchester}}]{2006MNRAS.372.1549E}
{Edwards}, R.~T., {Hobbs}, G.~B., \& {Manchester}, R.~N. 2006, \mnras, 372,
  1549

\bibitem[{{Foster} \& {Backer}(1990)}]{1990ApJ...361..300F}
{Foster}, R.~S., \& {Backer}, D.~C. 1990, \apj, 361, 300

\bibitem[{{Hobbs} {et~al.}(2014){Hobbs}, {Dai}, {Manchester}, {Shannon},
  {Kerr}, {Lee}, \& {Xu}}]{2014arXiv1407.0435H}
{Hobbs}, G., {Dai}, S., {Manchester}, R.~N., {et~al.} 2014, preprint,
  arXiv:1407.0435

\bibitem[{{Hobbs} {et~al.}(2010){Hobbs}, {Archibald}, {Arzoumanian}, {Backer},
  {Bailes}, {Bhat}, {Burgay}, {Burke-Spolaor}, {Champion}, {Cognard}, {Coles},
  {Cordes}, {Demorest}, {Desvignes}, {Ferdman}, {Finn}, {Freire}, {Gonzalez},
  {Hessels}, {Hotan}, {Janssen}, {Jenet}, {Jessner}, {Jordan}, {Kaspi},
  {Kramer}, {Kondratiev}, {Lazio}, {Lazaridis}, {Lee}, {Levin}, {Lommen},
  {Lorimer}, {Lynch}, {Lyne}, {Manchester}, {McLaughlin}, {Nice}, {Oslowski},
  {Pilia}, {Possenti}, {Purver}, {Ransom}, {Reynolds}, {Sanidas}, {Sarkissian},
  {Sesana}, {Shannon}, {Siemens}, {Stairs}, {Stappers}, {Stinebring},
  {Theureau}, {van Haasteren}, {van Straten}, {Verbiest}, {Yardley}, \&
  {You}}]{2010CQGra..27h4013H}
{Hobbs}, G., {Archibald}, A., {Arzoumanian}, Z., {et~al.} 2010, Classical and
  Quantum Gravity, 27, 084013

\bibitem[{{Hotan} {et~al.}(2004{\natexlab{a}}){Hotan}, {Bailes}, \&
  {Ord}}]{2004MNRAS.355..941H}
{Hotan}, A.~W., {Bailes}, M., \& {Ord}, S.~M. 2004{\natexlab{a}}, \mnras, 355,
  941

\bibitem[{{Hotan} {et~al.}(2004{\natexlab{b}}){Hotan}, {van Straten}, \&
  {Manchester}}]{2004PASA...21..302H}
{Hotan}, A.~W., {van Straten}, W., \& {Manchester}, R.~N. 2004{\natexlab{b}},
  \pasa, 21, 302

\bibitem[{{Janssen} {et~al.}(2015){Janssen}, {Hobbs}, {McLaughlin}, {Bassa},
  {Deller}, {Kramer}, {Lee}, {Mingarelli}, {Rosado}, {Sanidas}, {Sesana},
  {Shao}, {Stairs}, {Stappers}, \& {Verbiest}}]{2015aska.confE..37J}
{Janssen}, G., {Hobbs}, G., {McLaughlin}, M., {et~al.} 2015, Advancing
  Astrophysics with the Square Kilometre Array (AASKA14), 37

\bibitem[{{Karastergiou} {et~al.}(2011){Karastergiou}, {Roberts}, {Johnston},
  {Lee}, {Weltevrede}, \& {Kramer}}]{2011MNRAS.415..251K}
{Karastergiou}, A., {Roberts}, S.~J., {Johnston}, S., {et~al.} 2011, \mnras,
  415, 251

\bibitem[{{Kramer} {et~al.}(1999){Kramer}, {Xilouris}, {Camilo}, {Nice},
  {Backer}, {Lange}, {Lorimer}, {Doroshenko}, \&
  {Sallmen}}]{1999ApJ...520..324K}
{Kramer}, M., {Xilouris}, K.~M., {Camilo}, F., {et~al.} 1999, \apj, 520, 324

\bibitem[{{Kramer} {et~al.}(2006){Kramer}, {Stairs}, {Manchester},
  {McLaughlin}, {Lyne}, {Ferdman}, {Burgay}, {Lorimer}, {Possenti}, {D'Amico},
  {Sarkissian}, {Hobbs}, {Reynolds}, {Freire}, \&
  {Camilo}}]{2006Sci...314...97K}
{Kramer}, M., {Stairs}, I.~H., {Manchester}, R.~N., {et~al.} 2006, Science,
  314, 97

\bibitem[{{Lentati} {et~al.}(2015){Lentati}, {Alexander}, \&
  {Hobson}}]{2015MNRAS.447.2159L}
{Lentati}, L., {Alexander}, P., \& {Hobson}, M.~P. 2015, \mnras, 447, 2159

\bibitem[{{Lentati} {et~al.}(2014){Lentati}, {Alexander}, {Hobson}, {Feroz},
  {van Haasteren}, {Lee}, \& {Shannon}}]{2014MNRAS.437.3004L}
{Lentati}, L., {Alexander}, P., {Hobson}, M.~P., {et~al.} 2014, \mnras, 437,
  3004

\bibitem[{{Lentati} \& {Shannon}(2015)}]{2015MNRAS.454.1058L}
{Lentati}, L., \& {Shannon}, R.~M. 2015, \mnras, 454, 1058

\bibitem[{{Lentati} {et~al.}(2016){Lentati}, {Shannon}, {Coles}, {Verbiest},
  {van Haasteren}, {Ellis}, {Caballero}, {Manchester}, {Arzoumanian}, {Babak},
  {Bassa}, {Bhat}, {Brem}, {Burgay}, {Burke-Spolaor}, {Champion}, {Chatterjee},
  {Cognard}, {Cordes}, {Dai}, {Demorest}, {Desvignes}, {Dolch}, {Ferdman},
  {Fonseca}, {Gair}, {Gonzalez}, {Graikou}, {Guillemot}, {Hessels}, {Hobbs},
  {Janssen}, {Jones}, {Karuppusamy}, {Keith}, {Kerr}, {Kramer}, {Lam}, {Lasky},
  {Lassus}, {Lazarus}, {Lazio}, {Lee}, {Levin}, {Liu}, {Lynch}, {Madison},
  {McKee}, {McLaughlin}, {McWilliams}, {Mingarelli}, {Nice}, {Os{\l}owski},
  {Pennucci}, {Perera}, {Perrodin}, {Petiteau}, {Possenti}, {Ransom},
  {Reardon}, {Rosado}, {Sanidas}, {Sesana}, {Shaifullah}, {Siemens}, {Smits},
  {Stairs}, {Stappers}, {Stinebring}, {Stovall}, {Swiggum}, {Taylor},
  {Theureau}, {Tiburzi}, {Toomey}, {Vallisneri}, {van Straten}, {Vecchio},
  {Wang}, {Wang}, {You}, {Zhu}, \& {Zhu}}]{iptanoise}
{Lentati}, L., {Shannon}, R.~M., {Coles}, W.~A., {et~al.} 2016, \mnras, 458,
  2161

\bibitem[{{Lyne} {et~al.}(2010){Lyne}, {Hobbs}, {Kramer}, {Stairs}, \&
  {Stappers}}]{2010Sci...329..408L}
{Lyne}, A., {Hobbs}, G., {Kramer}, M., {Stairs}, I., \& {Stappers}, B. 2010,
  Science, 329, 408

\bibitem[{{Manchester} {et~al.}(2013){Manchester}, {Hobbs}, {Bailes}, {Coles},
  {van Straten}, {Keith}, {Shannon}, {Bhat}, {Brown}, {Burke-Spolaor},
  {Champion}, {Chaudhary}, {Edwards}, {Hampson}, {Hotan}, {Jameson}, {Jenet},
  {Kesteven}, {Khoo}, {Kocz}, {Maciesiak}, {Oslowski}, {Ravi}, {Reynolds},
  {Sarkissian}, {Verbiest}, {Wen}, {Wilson}, {Yardley}, {Yan}, \&
  {You}}]{2013PASA...30...17M}
{Manchester}, R.~N., {Hobbs}, G., {Bailes}, M., {et~al.} 2013, PASA, 30, 17

\bibitem[{{Nan} {et~al.}(2011){Nan}, {Li}, {Jin}, {Wang}, {Zhu}, {Zhu},
  {Zhang}, {Yue}, \& {Qian}}]{2011IJMPD..20..989N}
{Nan}, R., {Li}, D., {Jin}, C., {et~al.} 2011, International Journal of Modern
  Physics D, 20, 989

\bibitem[{{Ramachandran} \& {Kramer}(2003)}]{2003AA...407.1085R}
{Ramachandran}, R., \& {Kramer}, M. 2003, \aap, 407, 1085

\bibitem[{{Reardon} {et~al.}(2016){Reardon}, {Hobbs}, {Coles}, {Levin},
  {Keith}, {Bailes}, {Bhat}, {Burke-Spolaor}, {Dai}, {Kerr}, {Lasky},
  {Manchester}, {Os{\l}owski}, {Ravi}, {Shannon}, {van Straten}, {Toomey},
  {Wang}, {Wen}, {You}, \& {Zhu}}]{2016MNRAS.455.1751R}
{Reardon}, D.~J., {Hobbs}, G., {Coles}, W., {et~al.} 2016, \mnras, 455, 1751

\bibitem[{{Refregier}(2003)}]{2003MNRAS.338...35R}
{Refregier}, A. 2003, \mnras, 338, 35

\bibitem[{{Rickett}(1990)}]{1990ARAA..28..561R}
{Rickett}, B.~J. 1990, \araa, 28, 561

\bibitem[{{Shannon} \& {Cordes}(2010)}]{sc2010}
{Shannon}, R.~M., \& {Cordes}, J.~M. 2010, \apj, 725, 1607

\bibitem[{{Shannon} {et~al.}(2013){Shannon}, {Cordes}, {Metcalfe}, {Lazio},
  {Cognard}, {Desvignes}, {Janssen}, {Jessner}, {Kramer}, {Lazaridis},
  {Purver}, {Stappers}, \& {Theureau}}]{2013ApJ...766....5S}
{Shannon}, R.~M., {Cordes}, J.~M., {Metcalfe}, T.~S., {et~al.} 2013, \apj, 766,
  5

\bibitem[{{Shannon} {et~al.}(2015){Shannon}, {Ravi}, {Lentati}, {Lasky},
  {Hobbs}, {Kerr}, {Manchester}, {Coles}, {Levin}, {Bailes}, {Bhat},
  {Burke-Spolaor}, {Dai}, {Keith}, {Os{\l}owski}, {Reardon}, {van Straten},
  {Toomey}, {Wang}, {Wen}, {Wyithe}, \& {Zhu}}]{2015Sci...349..1522S}
{Shannon}, R.~M., {Ravi}, V., {Lentati}, L.~T., {et~al.} 2015, Science, 349,
  1522

\bibitem[{{Shao} {et~al.}(2013){Shao}, {Caballero}, {Kramer}, {Wex},
  {Champion}, \& {Jessner}}]{2013CQGra..30p5019S}
{Shao}, L., {Caballero}, R.~N., {Kramer}, M., {et~al.} 2013, Classical and
  Quantum Gravity, 30, 165019

\bibitem[{{Simonetti} {et~al.}(1984){Simonetti}, {Cordes}, \&
  {Spangler}}]{1984ApJ...284..126S}
{Simonetti}, J.~H., {Cordes}, J.~M., \& {Spangler}, S.~R. 1984, \apj, 284, 126

\bibitem[{{Tiburzi} {et~al.}(2016){Tiburzi}, {Hobbs}, {Kerr}, {Coles}, {Dai},
  {Manchester}, {Possenti}, {Shannon}, \& {You}}]{2016MNRAS.455.4339T}
{Tiburzi}, C., {Hobbs}, G., {Kerr}, M., {et~al.} 2016, \mnras, 455, 4339

\bibitem[{{van Straten}(2013)}]{2013ApJS..204...13V}
{van Straten}, W. 2013, \apjs, 204, 13

\bibitem[Weisberg et al.(1989)]{1989ApJ...347.1030W} Weisberg, J.~M., Romani, R.~W., \& Taylor, J.~H.\ 1989, \apj, 347, 1030 


\bibitem[{{Yu} {et~al.}(2013){Yu}, {Manchester}, {Hobbs}, {Johnston}, {Kaspi},
  {Keith}, {Lyne}, {Qiao}, {Ravi}, {Sarkissian}, {Shannon}, \&
  {Xu}}]{2013MNRAS.429..688Y}
{Yu}, M., {Manchester}, R.~N., {Hobbs}, G., {et~al.} 2013, \mnras, 429, 688

\end{thebibliography}

\end{document}